\documentclass[reprint,superscriptaddress]{revtex4-1}
\usepackage{amsmath,graphicx}
\begin{document}

\title{Speckle-based determination of the polarisation state of single \\ and multiple laser beams}

\author{Morgan Facchin} \email{mf225@st-andrews.ac.uk}
\affiliation{SUPA School of Physics and Astronomy, University of St Andrews, North Haugh, St Andrews KY16 9SS, UK}

\author{Graham D. Bruce}
\affiliation{SUPA School of Physics and Astronomy, University of St Andrews, North Haugh, St Andrews KY16 9SS, UK}
\author{Kishan Dholakia}
\affiliation{SUPA School of Physics and Astronomy, University of St Andrews, North Haugh, St Andrews KY16 9SS, UK}
\affiliation{Department of Physics, College of Science, Yonsei University, Seoul 03722, South Korea}
\affiliation{Graduate School of Engineering, Chiba University, 1-33 Yayoi-cho, Inage-ku, Chiba-shi 263-0022, Japan}

\begin{abstract}
Laser speckle is generated by the multiple interference of light through a disordered medium. Here we study the premise that the speckle pattern retains information about the polarisation state of the incident field. We analytically verify that a linear relation exists between the Stokes vector of the light and the resulting speckle pattern. As a result, the polarisation state of a beam can be measured from the speckle pattern using a transmission matrix approach. We perform a quantitative analysis of the accuracy of the transmission matrix method to measure randomly time-varying polarisation states. In experiment, we find that the Stokes parameters of light from a diode laser can be retrieved with an uncertainty of 0.05 using speckle images of 150$\times$150 pixels and 17 training states. We show both analytically and in experiment that this approach may be extended to the case of more than one laser field, demonstrating the measurement of the Stokes parameters of two laser beams simultaneously from a single speckle pattern and achieving the same uncertainty of 0.05. 
\end{abstract}

\maketitle
\section{Introduction}
When coherent light is diffused by a disordered medium, it produces a typical granular pattern called speckle. Despite their random and uncontrollable nature, speckle patterns encode information about both the diffuser and the light, and can therefore be used to perform a range of measurements \cite{Goodman}. Two approaches are possible: if one considers the incident light to be time-invariant, the speckle pattern can be harnessed to probe properties of the diffuser. This is the dominant idea in speckle metrology. This has been applied to many types of measurements, such as displacement \cite{Archbold70,freund1990surface,burch1968,wang2006core}, vibration and sound \cite{bianchi14,zalevsky09}, blood flow mapping in tissues \cite{briers13}, among many others. Another approach is to consider the diffuser to be constant in time, in which case the speckle pattern can be harnessed to probe properties of the incident light. This more recent concept has been applied to the measurement of wavelength variations and laser stabilisation \cite{Mazilu14,hanson2015speckle,chakrabarti2015speckle,Metzger17,bruce19,Bruce20,odonnellhigh,Gupta19}, spectroscopy \cite{redding2012using,redding2013compact,redding2013all,Cao17}, and transverse mode characterisation of structured light \cite{Mourka13,Mazilu12}. 

Measurement of polarisation of a light field is a key requirement in a breadth of photonics applications, and studies related to polarisation measurement with speckle have also been carried out. A previous study derived an explicit expression for estimating the Stokes parameters of a laser beam, in terms of the cross-correlations between four particular speckle patterns corresponding to the four classical polarisation filters \cite{freund90}. Later studies included a generalised approach using Jones-like transmission matrices \cite{kohlgraf2008transfer}, and Mueller-like transmission matrices allowing spatially resolved polarimetry \cite{kohlgraf2009spatially}, and spectropolarimetry \cite{kohlgraf2010transmission}. 

In this paper we focus on quantitative analysis of the transmission matrix method performance when applied to polarisation measurement. The measurements are performed on randomly time-varying polarisation states of the incident field. Importantly we also extend the method to the case of multiple beams, where the polarisation state of multiple beams can be measured simultaneously from one single speckle pattern. We demonstrate this for the case of two light fields. We also provide a demonstration verifying the linearity between the Stokes vector of the input beam and the resulting speckle pattern, and extend this result to the case of multiple beams. Such an approach may have applications in optical telecommunications, optical manipulation of birefringent particles and polarisation microscopy.

\section{Background}
\label{sec:examples}
Changing the polarisation state of a laser beam, for example by manually rotating a waveplate, induces a visible change in the speckle pattern produced after diffusion. 
In this section we derive an expression for the linearity that exists between the polarisation state and the speckle pattern, which is at the core of the transmission matrix method.  

\begin{figure}[h!]
\centering\includegraphics[width=\linewidth]{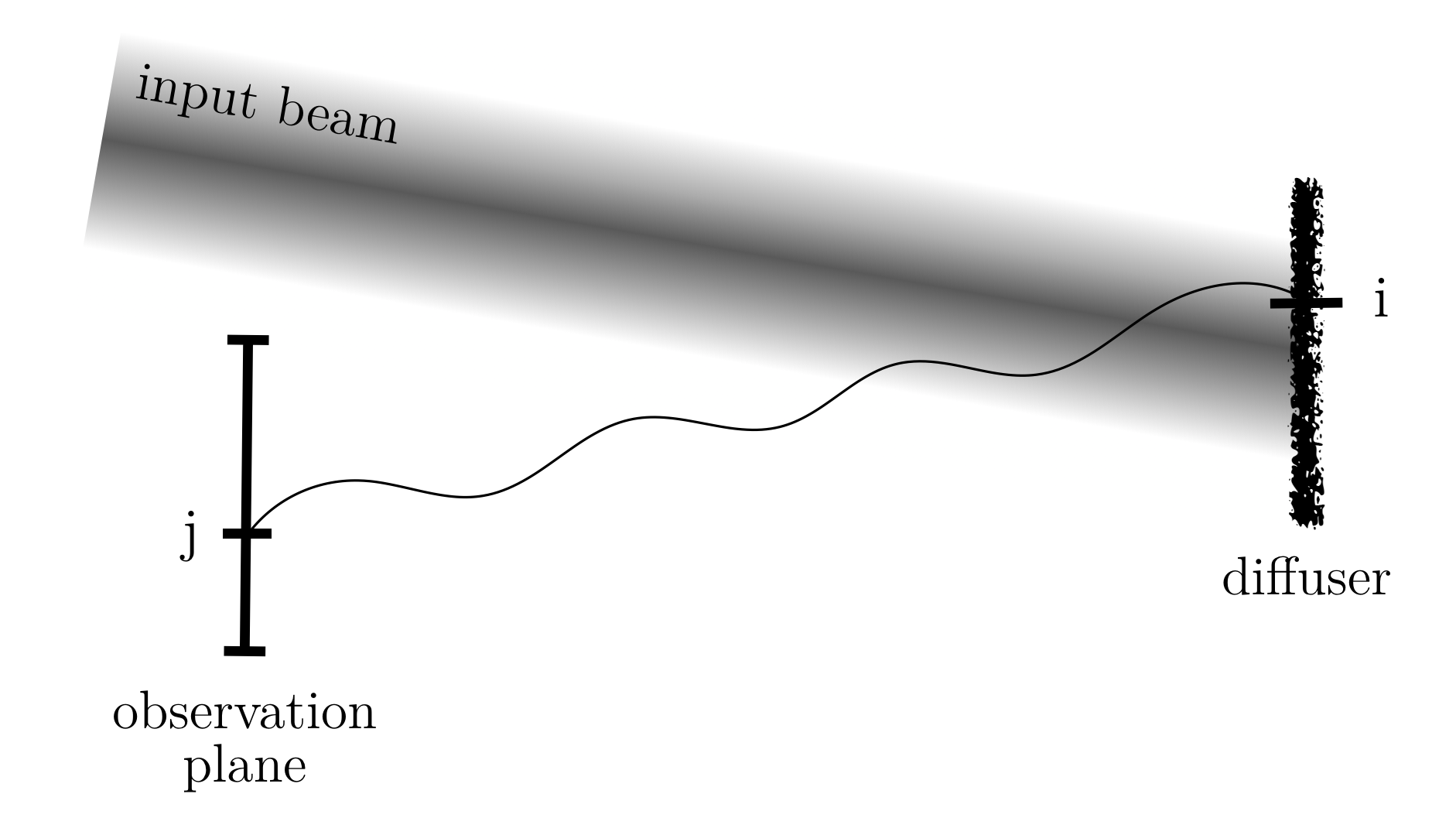}
\caption{\textbf{Diffusion geometry.} An input laser beam is incident on a single rough surface, which diffuses the light in the same half-space. The diffused light is collected on a surface denoted as observation plane. }
\label{fig:geom}
\end{figure}

We consider the geometry displayed in figure \ref{fig:geom} where an input beam of spatially constant polarisation state is diffused by a single reflective surface, which we model as an assembly of discrete elements. The light wave is described by its electric component $\bf{E}$, which is a $1\times3$ complex vector. We assume that the electric field incident upon point $j$ of the observation plane after being diffused by the $i$th element of the diffuser is of the form
\begin{equation} 
{\bf{E}}_{ij}={\bf{E}}_{i} \alpha _{ij},
\end{equation}
where ${\bf{E}}_{i}$ is the electric field at point $i$, and $\alpha _{ij}$ is a complex tensor containing all the details of the field's transformation from $i$ to $j$, which includes the contribution of the diffusion itself, as well the transport from $i$ to $j$ \cite{kohlgraf2008transfer}. This relation expresses the linear nature of the diffusion process, which is our assumption. 
The total field at $j$ is the sum of the fields propagating from all the illuminated points of the diffuser, which gives
\begin{equation} 
{\bf{E}}_{j}=\sum_{i}{\bf{E}}_{i}\alpha _{ij}.
\end{equation}
As the input beam has a spatially constant polarisation state, the field incident upon $i$ can be written as ${\bf{E}}_{i}={\bf{e}}\rho_{i}e^{i\phi_{i}+i\varphi (t)}$, where $\rho_{i}$ is the amplitude of the electric field at point $i$, $\phi_{i}$ is the spatial part of the phase at point $i$, $\varphi (t)$ is the temporal part of the phase (containing the $\omega t$ term and any kind of temporal fluctuations, common to all $i$), and ${\bf{e}}$ is the normalised Jones vector. The resultant field at $j$ is then

\begin{equation} \label{eq:alpha}
\begin{split}
{\bf{E}}_{j}&=\sum_{i}{\bf{e}}\,\rho_{i}e^{i\phi_{i}+i\varphi (t)}\alpha _{ij} \\
&={\bf{e}}\left ( \sum_{i}\rho_{i}e^{i\phi_{i}}\alpha _{ij} \right ) e^{i\varphi (t)}\\
&={\bf{e}} \alpha _{j} e^{i\varphi (t)},\\
\end{split}
\end{equation}
where $\alpha_{j}$ contains all the details of the field's transformation terminating at point $j$.

It is established that when the field undergoes a linear transformation, as the last relation indicates, the Stokes vector also undergoes a linear transformation \cite{kim1987relationship}. We show later in this paper that we can extend this to the case of multiple beams.   
First, we express the coherency matrix, which is defined as 
$C= \langle \bf{E}\otimes\bf{E}^{*} \rangle$, 
where the brackets denote time averaging, $\otimes$ the outer (or Kronecker) product, and $*$ the complex conjugate. Using the fact that the outer product can be expressed as a usual matrix product when the vectors are properly shaped, the coherency matrix at point $j$ is  
\begin{equation} 
\begin{split}
 C_{j}&=\left \langle{\bf{E}}_{j} \otimes {\bf{E}}_{j}^{*}\right \rangle\\
      &=\left \langle{\bf{E}}_{j}^{'} {\bf{E}}_{j}^{*}\right \rangle \\
      &=\left \langle \left ( {\bf{e}}\alpha_{j}e^{i\varphi (t)}  \right )^{'} \left ({\bf{e}}\alpha_{j}e^{i\varphi (t)}\right )^{*}\right \rangle \\
      &=(\alpha_{j}^{'}{\bf{e}}^{'}) ({\bf{e}}^{*}\alpha_{j}^{*}) \\
      &=\alpha_{j}^{'}({\bf{e}}^{'} {\bf{e}}^{*})\alpha_{j}^{*} \\
      &=\alpha _{j}^{'}({\bf{e}} \otimes {\bf{e}}^{*})\alpha _{j}^{*} \\
      &=\alpha _{j}^{'}C_{0}\alpha _{j}^{*}, \\
\end{split}
\end{equation}
with the prime denoting transposition, and $C_{0}$ the coherency matrix of the input beam normalised to the intensity (as it is computed from the normalised Jones vector). Note that, by definition, the electric field in the expression of the coherency matrix is no longer a $1\times3$ vector, but a $1\times2$ vector, expressed in the plane perpendicular to the direction of propagation. As this direction is arbitrary (the observation plane can be anywhere), a change of basis in which the electric field is expressed is required. The change of coordinate, as well as the cropping of the electric field, can be performed by a matrix multiplication of ${\bf{E}}_{j}$, which we implicitly absorb in $\alpha _{j}$ for clarity. 

Now that we know how the coherency matrix transforms, we wish to determine how the Stokes parameters subsequently transform. 
The Stokes parameters are related to the coherency matrix via the relation $S^{m}=Tr(C\sigma_{m})$, where $Tr$ is the trace, $S^{m}$ is the $m$th Stokes parameter, and $\sigma_{m}$ is the $m$th Pauli matrix. This operation is analogous to a projection of the coherency matrix onto the three Pauli matrices (to which is added the unit matrix), as the Stokes parameters are defined as twice the coefficients of decomposition of the coherency matrix in this basis, expressed as $C=\sum_{n}\frac{1}{2}S^{n}\sigma_{n}$ \cite{kim1987relationship}.
Applying these two relations at point $j$ we have

\begin{equation} 
\begin{split}
S_{j}^{m}&=Tr(C_{j}\sigma_{m}) \\
&=Tr(\alpha _{j}^{'}C_{0}\alpha _{j}^{*}\sigma_{m})\\
&=Tr(\alpha _{j}^{'}\sum_{n}\frac{1}{2}S^{n}\sigma_{n}\alpha _{j}^{*}\sigma_{m})\\
&=\sum_{n}S^{n}\frac{1}{2}Tr(\alpha _{j}^{'}\sigma_{n}\alpha _{j}^{*}\sigma_{m})\\
{\bf{S}}_{j}&={\bf{S}}M_{j}. \\
\end{split}
\end{equation}

Finally, we find that the Stokes vector at point $j$ is linearly related to the Stokes vector of the input beam ${\bf{S}}$, through a $4\times4$ matrix $M_{j}$, the elements of which are given by $M_{j,nm}=\frac{1}{2}Tr(\alpha _{j}^{'}\sigma_{n}\alpha _{j}^{*}\sigma_{m})$. In other words, $M_{j}$ is the Mueller matrix associated to the diffuser at point $j$. Note that ${\bf{S}}$ is the normalised Stokes vector, as it is computed from the normalised coherency matrix.

\noindent As a camera only records the intensity, given by the first Stokes parameter $S_{j}^{0}$, only the first column of $M_{j}$ is needed, and the intensity observed at point $j$ is given by $I_{j}={\bf{S}}{\bf{M}}_{j,1}$, with ${\bf{M}}_{j,1}$ the first column of $M_{j}$.

\section{Method}
The last relation found above can be extended to any set of points on the observation plane, and leads to the following central relation 
\begin{equation}
{\bf{I}}={\bf{S}}M,
\label{eq:lin}
\end{equation}
with $\bf{I}$ the $1\times L$ image of the speckle pattern reshaped into a row vector, and $\bf{S}$ the $1\times 4$ Stokes vector of the input beam. $M$ is a $4\times L$ matrix making the connection between the two, and is usually referred to as transmission (or transfer, or measurement) matrix \cite{popoff2010measur_transm,kohlgraf2010transmission,kohlgraf2009spatially}. 

The transmission matrix is unknown and depends on many parameters, such as the spectrum, beam profile, and angle of incidence of the input beam, its position on the diffuser, and the detailed structure of the diffuser. However, assuming all those conditions to be time-invariant, we can determine the transmission matrix using a set of $N$ (with any $N\geq 4$) training polarisation states and their corresponding speckle patterns. Applying (\ref{eq:lin}) to the training sets and performing a simple matrix inversion leads to  
\begin{equation}
I_{0}=S_{0}M,
\label{eq:system}
\end{equation}
\begin{equation}
M=S_{0}^{+}I_{0},
\label{eq:solve}
\end{equation}

with $I_{0}$ a $N\times L$ matrix containing the training speckle patterns stacked in rows, and $S_{0}$ a $N\times 4$ matrix containing the corresponding Stokes parameters ($S_{0}^+$ being the $4\times N$ pseudo-inverse of $S_{0}$). Here the pseudo-inverse is used, rather than the standard inverse, because $S_{0}$ is not square. 
Indeed, four training states would be enough to close the system, but in practice better results are obtained using more training states (see section 4 and 5), in which case $S_{0}$ is rectangular. With $N> 4$, expression (\ref{eq:system}) is an over-determined system, and solving by (\ref{eq:solve}) corresponds to the minimisation of the Euclidean (or Frobenius, or L2) norm $\left \| I_{0}-S_{0}M \right \|^{2}$, defined by $\left \|  A\right \|^{2}=\sum_{ij}A_{ij}^{2}$. 

Armed with $M$, we now have all we need in (\ref{eq:lin}) to find the Stokes parameters of any new polarisation state, given its corresponding speckle pattern. By solving for the Stokes vector we find 
\begin{equation}
{\bf{S}}={\bf{I}} M^{+}.
\label{eq:S}
\end{equation}
As a passing remark, one can see from this last relation that what is needed in practice is $M^{+}$ and not $M$. To avoid unnecessary interim calculations, one can directly compute $M^{+}$ from the training sets given by $M^{+}=I_{0}^{+}S_{0}$. 

\section{Experimental implementation}
\begin{figure}[h!]
\centering\includegraphics[width=\linewidth]{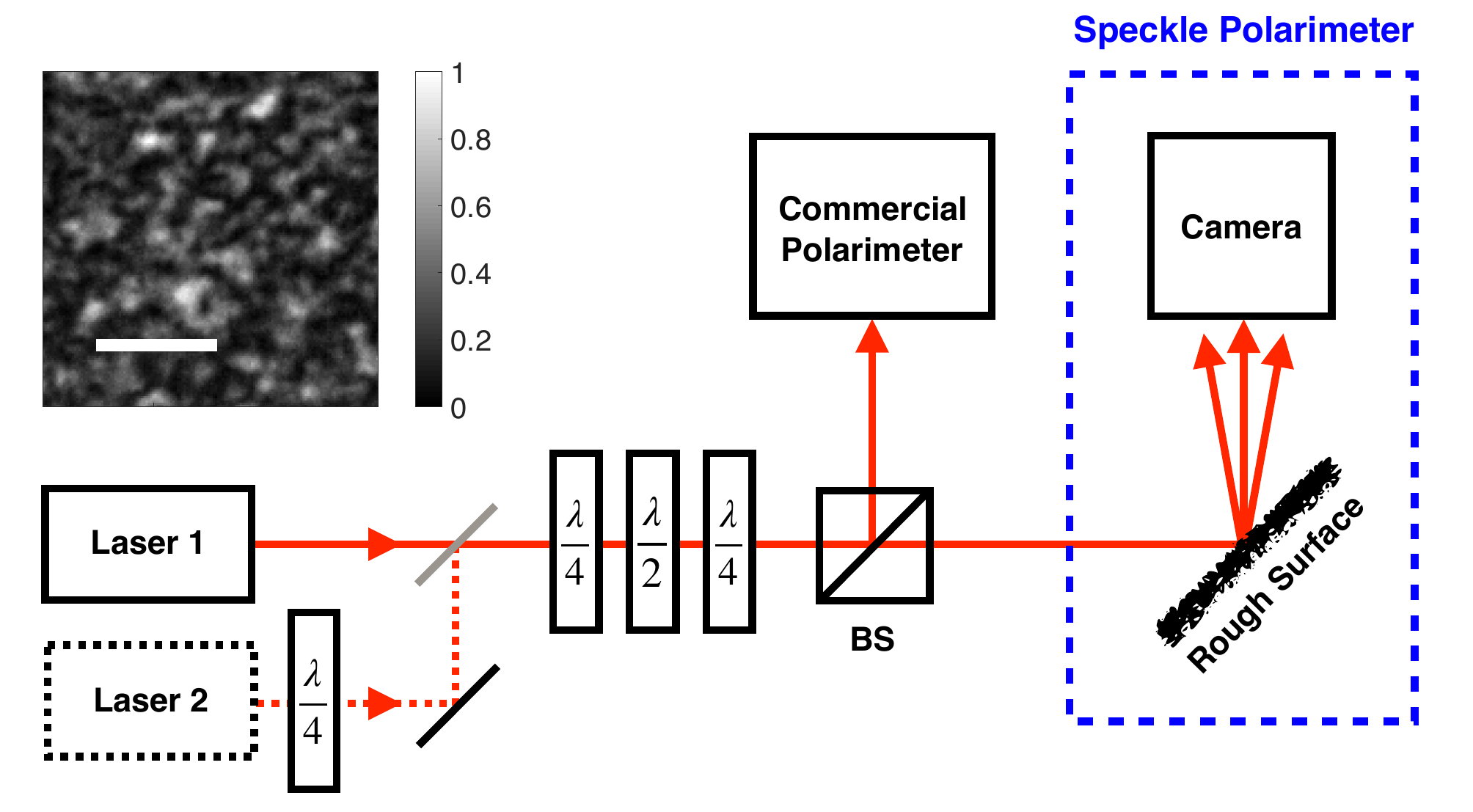}
\caption{Polarisation measurement setup. A laser beam passes through three waveplates, rotating with incommensurate angular speeds, enabling a randomly time-varying state of polarisation. The light is split into two paths using a non-polarising beam splitter (BS): on one path the state of polarisation is measured using a commercial polarimeter, on the other path the laser is diffused on a single high-reflective rough surface, and the produced speckle pattern is recorded on a CMOS camera. A $150\times150$-pixels image of a speckle pattern is shown, with each pixel being 8 $\mu$m $\times$ 8 $\mu$m. The scale bar denotes 50 pixels (0.4 mm), and the colour bar shows the intensity normalised to maximum. For the multiple (two-beams) version of the experiment (see section 5), a second laser joins the optical path via a pellicle beam splitter (gray), after passing through a waveplate so that the state of polarisation of laser 2 is different to that of laser 1 before passing through the three waveplates. }
\label{fig:setup}
\end{figure}

Our setup is shown in figure \ref{fig:setup}. A laser beam of $780$ nm wavelength (7 mW power and 1 MHz linewidth) is initially set to a highly stable linear polarisation state by means of a 40 dB optical isolator (Isowave I-80-T-5-H). It then passes through three successive waveplates (respectively quarter wave, half wave, and quarter wave), mounted on independent motorised rotating stages (Thorlabs KPRM1E/M), to allow generation of an arbitrary and dynamically controlled polarisation state. The beam is then split into two paths using a non-polarising beam splitter. One path leads to a commercial polarimeter (Thorlabs PAX1000IR1/M), and one path leads to a rough surface where the light impinges at 45$^{\circ}$ and is diffused to form a speckle pattern directly on the camera (Mikrotron MotionBLITZ EoSens mini2) without intermediate lenses. The rough surface is a 12.5 mm-diameter, 1 mm-thick disk of a PTFE-based material with high reflectivity and highly Lambertian reflectance in the 250 - 2500 nm wavelength range (Thorlabs SM05CP2C). The distance between rough surface and camera is chosen to be 12 cm, so that individual grains in the speckle pattern cover approximately 15$\times$15 pixels. We simultaneously record the polarisation measured by the commercial polarimeter and the speckle pattern. An example of an obtained speckle pattern is shown in figure \ref{fig:setup}. 

\begin{figure}[h!]
\centering\includegraphics[width=\linewidth]{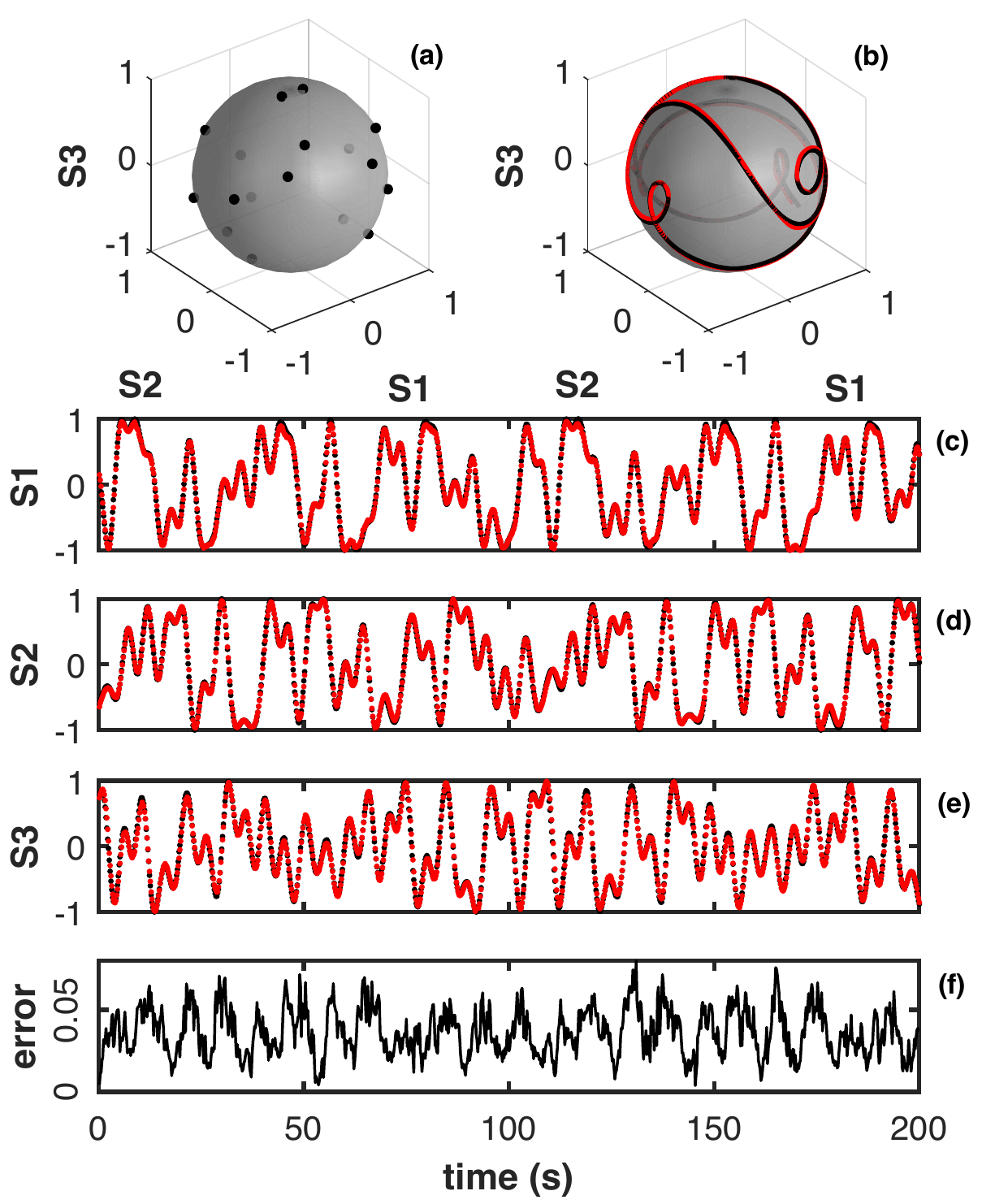}
\caption{\textbf{Single-beam polarisation measurement.} (a) Poincar\'e sphere representation of the 17 training states. (b) Trajectory of the polarisation state across the Poincar\'e sphere from t=160 s to t=190 s, measured by the commercial polarimeter (black) and retrieved from the speckle patterns (red). (c-e) The Stokes parameters $S_{1}$ to $S_{3}$ as a function of time, measured by the commercial polarimeter (black) and retrieved from the speckle patterns (red). (f) The error as the absolute residual, was averaged over the Stokes parameters. The estimation was performed using 150$\times$150-pixels images and 17 training states. }
\label{fig:results}
\end{figure}

\begin{figure}[h!]
\centering\includegraphics[width=\linewidth]{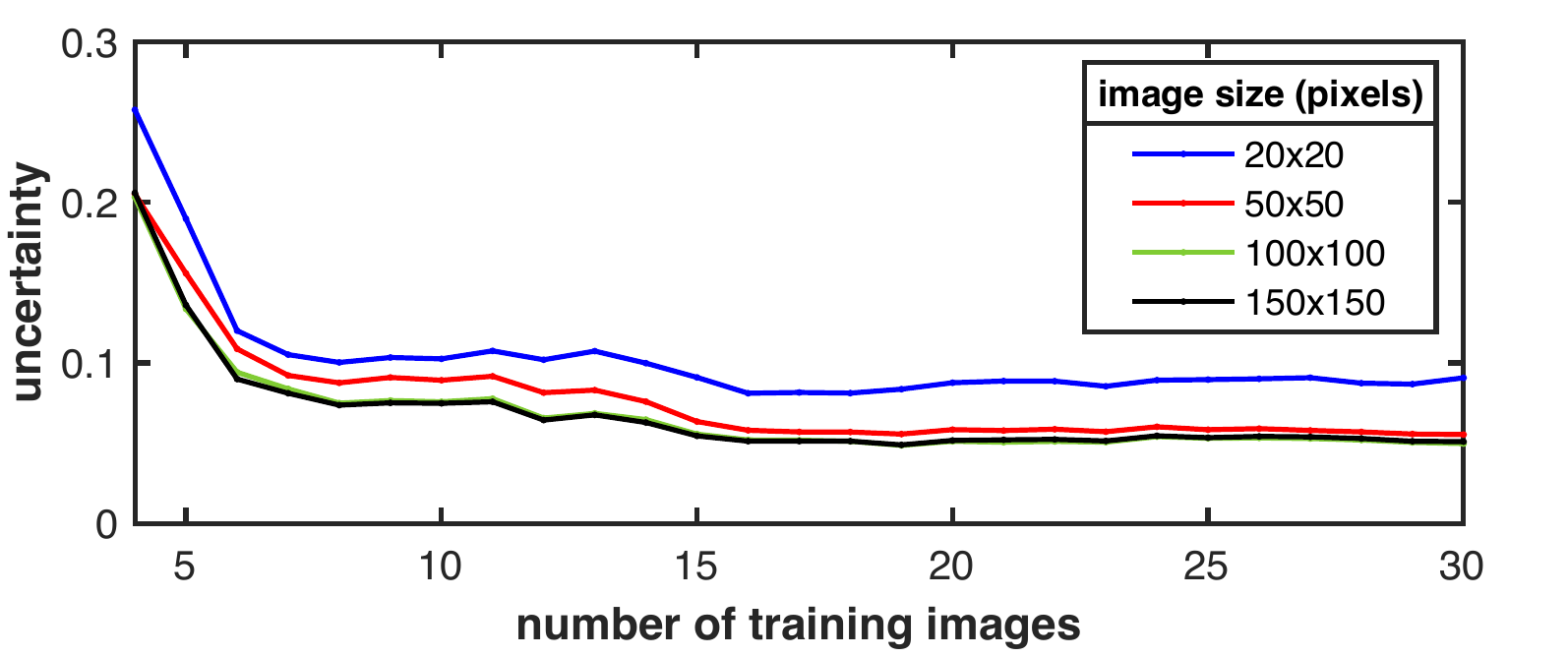}
\caption{\textbf{Measurement uncertainty for a single beam.} The uncertainty is given by the standard deviation of the residuals. It is shown as a function of the number of training images (four being the minimum required), for different image sizes ranging from $20\times20$ to $150\times150$ pixels. We see that the uncertainty reaches a minimum of 0.05 after about 15 training images and an image size of 100$\times$100 pixels. }
\label{fig:uncertainty}
\end{figure}

In order to test the method described above, we prepared the laser beam in a random and continuously time-varying state of polarisation by rotating the waveplates with incommensurate angular speeds ($5.77^{\circ}s^{-1}$, $10.77^{\circ}s^{-1}$, and $20.77^{\circ}s^{-1}$). As the plates rotated, we simultaneously recorded the measurements of the commercial polarimeter and 150$\times$150-pixels images of the speckle patterns, at regular intervals of 0.2 s. We picked 17 states at the beginning of the time series to make up our training sets $S_{0}$ and $I_{0}$ respectively in (\ref{eq:system}). Once $M^{+}$ was determined, we estimated the Stokes parameters of the subsequent states by applying equation (\ref{eq:S}) to the speckle patterns and compared to the measurements of the commercial polarimeter. We show the results in figure \ref{fig:results}, and give a quantitative analysis of the uncertainty in terms of image size and number of training images in figure \ref{fig:uncertainty}. 

We define the uncertainty on the Stokes parameters retrieval as the standard deviation of the residuals, given by the difference between the measurements of the commercial polarimeter and the estimation from the speckle patterns. We find that the uncertainty rapidly reaches a value of 0.05 for about 15 training images and an image size of 100$\times$100 pixels. For comparison, the resolution of the commercial polarimeter is $0.01$. 

It is worth pointing out that what is retrieved from the speckle patterns is the polarisation of the light that is incident upon the polarimeter, not necessarily that of the light incident upon the diffuser. These may differ due to the reflection in the beam splitter, and are linearly related by the Mueller matrix associated with the reflection. For any unknown beam, the polarisation retrieved from the speckle patterns is the one that would be measured by the polarimeter. If this is of any importance in a given application, for example if one needs to determine the polarisation of the light incident upon the diffuser, one would simply have to determine the Mueller matrix of the beam splitter. That task would be strictly analogous to what is done in the method section, where the unknown matrix $M$ would be the Mueller matrix of the beam splitter, and the two linearly related vectors would be the polarisation states of the two paths.

\section{Multiplexing}
As the method relies on the acquisition of speckle images, it can be extended to the simultaneous measurement of multiple beams. Indeed, if multiple beams are diffused on the same surface, all the corresponding speckle patterns superpose on the camera. If, in addition, the different beams originate from different sources, then the speckle patterns superpose without interference. It follows that the polarisation state of each beam is linearly encoded in the resulting speckle pattern, analogously to the single-beam case, as the polarisation state of each beam is already linearly encoded in its own speckle pattern. In this section we show that a relation equivalent to (\ref{eq:lin}) holds in the case of two beams, allowing the same method to be applied, and we test it experimentally.

The electric field is now a superposition of two fields ${\bf{E}}={\bf{E}}_{1}+{\bf{E}}_{2}$. The electric field at point $j$ is then ${\bf{E}}_{j}={\bf{e}}_{1}\alpha _{1,j}e^{i\varphi_{1} (t)}+{\bf{e}}_{2}\alpha _{2,j}e^{i\varphi_{2} (t)}$, the alpha tensor being different for each beam, as they may hit different diffusers with different phases and amplitudes. Injecting this in the expression of the coherency matrix at point $j$ gives  

\begin{equation}
\begin{split}
 C_{j}= \; &\left \langle{\bf{E}}_{j} \otimes {\bf{E}}_{j}^{*} \right \rangle \\
 = \; &\bigg \langle \left ({\bf{e}}_{1}\alpha_{1,j}e^{i\varphi_{1} (t)}+{\bf{e}}_{2}\alpha_{2,j}e^{i\varphi_{2} (t)}
 \right )^{'} \times ...\\
 \; & \left ({\bf{e}}_{1}\alpha_{1,j}e^{i\varphi_{1} (t)}+{\bf{e}}_{2}\alpha_{2,j}e^{i\varphi_{2} (t)} \right )^{*} \bigg \rangle \\
  = \; &\alpha_{1,j}^{'}({\bf{e}}_{1}^{'} {\bf{e}}_{1}^{*})\alpha_{1,j}^{*} +
  \alpha_{2,j}^{'}({\bf{e}}_{2}^{'} {\bf{e}}_{2}^{*})\alpha_{2,j}^{*}\\
= \; &\alpha _{1,j}^{'}C_{1}\alpha _{1,j}^{*}+
\alpha _{2,j}^{'}C_{2}\alpha _{2,j}^{*},\\
\end{split}
\end{equation}
where the cross terms are zero as the beams are not coherent with each other, $C_{1}$ and $C_{2}$ are the normalised coherency matrices of each individual input beam. Now expressing the Stokes vector at point $j$ we have 

\begin{equation} \label{decomp2}
\begin{split}
S_{j}^{m}&=Tr(C_{j}\sigma_{m}) \\
&=Tr(\alpha _{j}^{'}C_{1}\alpha _{j}^{*}\sigma_{m}+\alpha _{j}^{'}C_{2}\alpha _{j}^{*}\sigma_{m})\\
&=Tr(\alpha _{j}^{'}C_{1}\alpha _{j}^{*}\sigma_{m})+Tr(\alpha _{j}^{'}C_{2}\alpha _{j}^{*}\sigma_{m})\\
{\bf{S}}_{j}&={\bf{S}}_{1}M_{1,j}+{\bf{S}}_{2}M_{2,j}, \\
\end{split}
\end{equation}
where $M_{1,j}$ and $M_{2,j}$ are the Mueller matrices associated to point $j$ for each beam, and ${\bf{S}}_{1}$ and ${\bf{S}}_{2}$ are the normalised Stokes vectors of each beam. Again, as only the intensity is observed on the camera, the intensity at point $j$ is given by ${\bf{I}}_{j}={\bf{S}}_{1}{\bf{M}}_{1,j,1}+{\bf{S}}_{2}{\bf{M}}_{2,j,1}$, where ${\bf{M}}_{1,j,1}$ and ${\bf{M}}_{2,j,1}$ are the first column of $M_{1,j}$ and $M_{2,j}$. ${\bf{I}}_{j}$ can actually be expressed in one single dot product as 
\begin{equation} 
{\bf{I}}_{j}=\Bar{{\bf{S}}}\Bar{{\bf{M}}}_{j},
\end{equation}
where the two Stokes vectors ${\bf{S}}_{1}$ and ${\bf{S}}_{2}$ are concatenated in one $1\times8$ vector $\Bar{{\bf{S}}}$, and $M_{1,j,1}$ and $M_{2,j,1}$ are concatenated in one $8\times1$ vector $\Bar{{\bf{M}}}_{j}$. Generalising again to a set of points on the observation plane, we finally find the two-beams equivalent of relation (\ref{eq:lin}):
\begin{equation} \label{eq:multi}
{\bf{I}}=\Bar{{\bf{S}}}\Bar{M},
\end{equation}
where ${\bf{I}}$ is the $1\times L$ speckle image, $\Bar{M}$ is the two-beams $8\times L$ transmission matrix. As this relation is of the same form as relation (\ref{eq:lin}), it implies that all the analysis performed in the method section can be applied in the exact same way. The only difference is that the inversion (\ref{eq:S}) gives both Stokes vectors in a single $1\times8$ vector. In principle this approach can be extended to more beams, a mathematical limit being one quarter of the number of pixels in the speckle image, and a physical limit being contrast reduction as the number of beams increases. 

\begin{figure}[h!]
\centering\includegraphics[width=\linewidth]{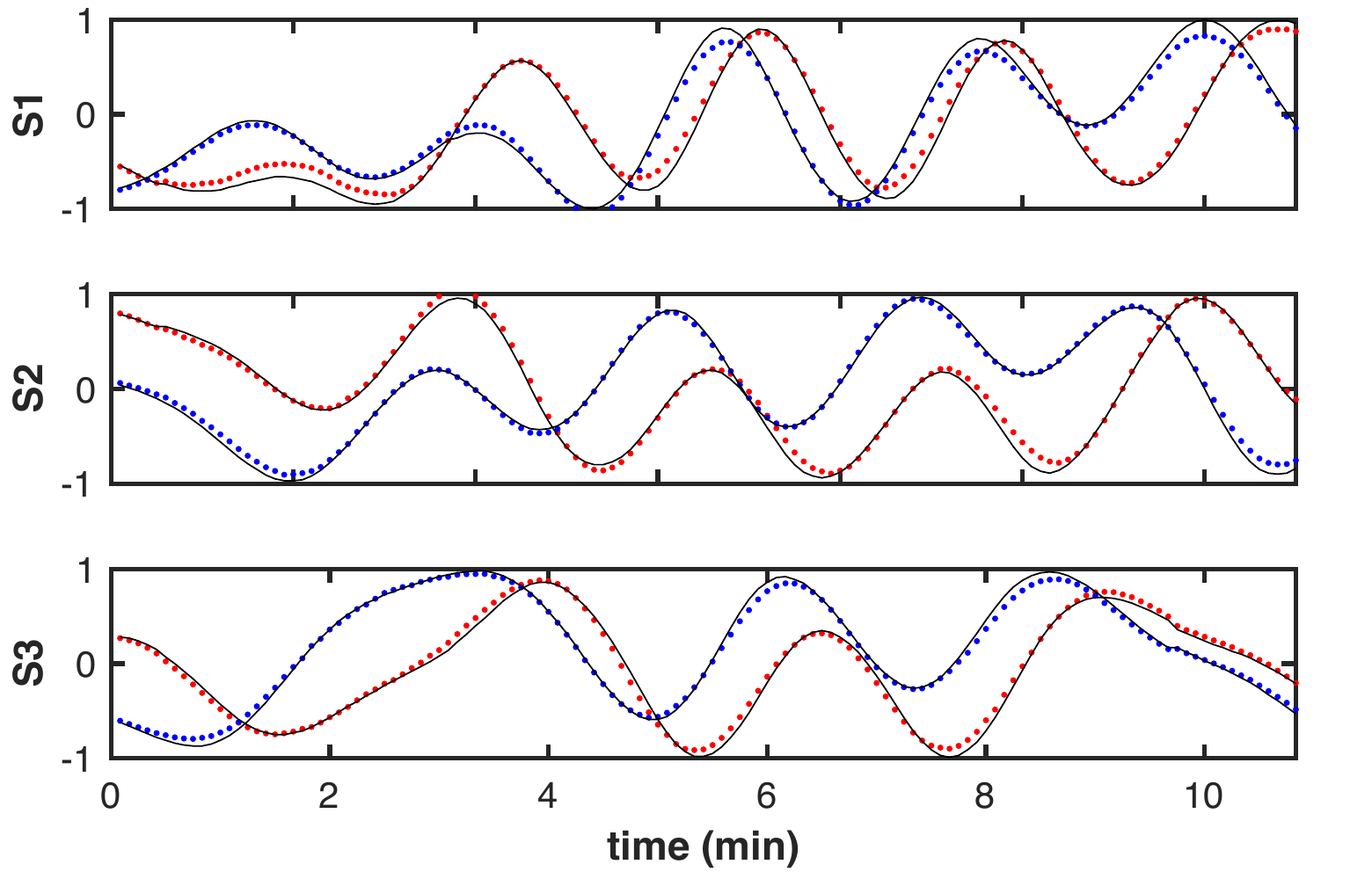}
\caption{\textbf{Two-beams polarisation measurement.} The Stokes parameters $S_{1}$ to $S_{3}$ are given as a function of time, measured by the commercial polarimeter (black) and retrieved from the speckle patterns (red and blue, one for each beam). The estimation was performed using 150$\times$150-pixels images and 17 training states.}
\label{fig:multi}
\end{figure}

\begin{figure}[h!]
\centering\includegraphics[width=\linewidth]{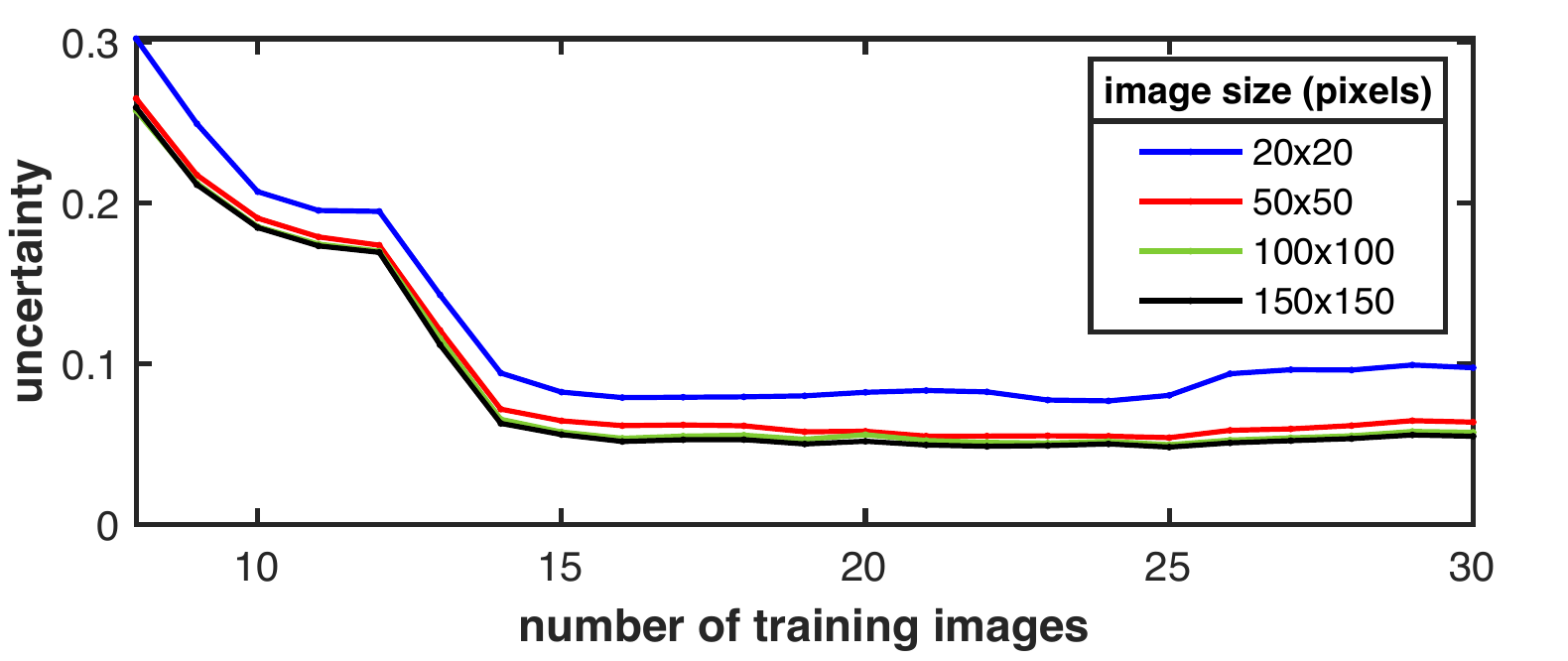}
\caption{\textbf{Measurement uncertainty for two beams.} The uncertainty is given by the standard deviation of the residuals. It is shown as a function of the number of training images (eight being the minimum required), for different image sizes ranging from $20\times20$ to $150\times150$ pixels. Here the uncertainty also reaches a minimum of 0.05 after about 15 training images.  }
\label{fig:uncertainty2}
\end{figure}

To test this relation, we added a second laser beam to the setup described in figure \ref{fig:setup}, of the same wavelength and power as the first one. This is also prepared in an initial high purity linear polarisation state by means of a 60 dB optical isolator (Toptica DSR780). We inserted a quarter waveplate in a fixed arbitrary orientation on the path of the second laser, so that its polarisation state was different than that of the first laser before entering the three rotating waveplates. This ensured that both beams hit the diffuser with different time-varying polarisation states. Also, we needed to know the polarisation state of each individual beam at any given time, which is not possible when they enter the commercial polarimeter simultaneously. Therefore we rotated the waveplates by small discrete increments, between which the waveplates were left stationary for 5 seconds. During those 5 seconds, three measurements were made: each beam was sequentially blocked while the polarisation state of the other one was measured by the commercial polarimeter, and an image of the speckle pattern was recorded when both beams hit the diffuser. This way of proceeding took an extended period of time, implying a timescale difference with our single-beam experiment and less data points. For consistency with the single-beam experiment, we also picked 17 training states at the beginning of the time series and applied the retrieval to the subsequent states, using 150$\times$150-pixels images. We show the results in figure \ref{fig:multi}, and the uncertainty analysis in \ref{fig:uncertainty2}. Interestingly, we find that the uncertainty reaches a minimum after approximately the same number of training images than in the single-beam case, i.e. 15 training images.

The method could be applied to beams of different and time-varying powers, without needing to train for different powers. In equation (\ref{eq:lin}) and (\ref{eq:multi}) we use the normalised Stokes vector, and the transmission matrix is a function of the power, respectively because in equation (\ref{eq:alpha}) $\bf{e}$ is normalised and $\alpha_{j}$ depends on the power. However if the training Stokes vectors are simply multiplied by the power, then the transmission matrix found by (\ref{eq:solve}) would be power-independent. It follows that the estimation (\ref{eq:S}) would give the non-normalised Stokes vector, for any input power. In the multiple-beams case, each training Stokes vector of $\Bar{{\bf{S}}}$ in (\ref{eq:multi}) would need to be individually multiplied by the power of its corresponding beam, and again the estimation would provide the non-normalised Stokes vectors, for any input powers.

\section{Acquisition speed and regularity }
Another advantage to our speckle-based polarisation measurement technique is that, when performed with a fast-framing camera, it allows a higher sampling rate compared to commercial polarimeters based on mechanically-rotating polarisers. The Thorlabs PAX1000IR1/M we use as a benchmark can theoretically achieve a sampling rate of 400 Hz, but in practice we found it limited to 110 Hz, with an irregular sampling. In this section we explore the high speed capability by applying faster polarisation changes. 

\begin{figure}[h!]
\centering\includegraphics[width=\linewidth]{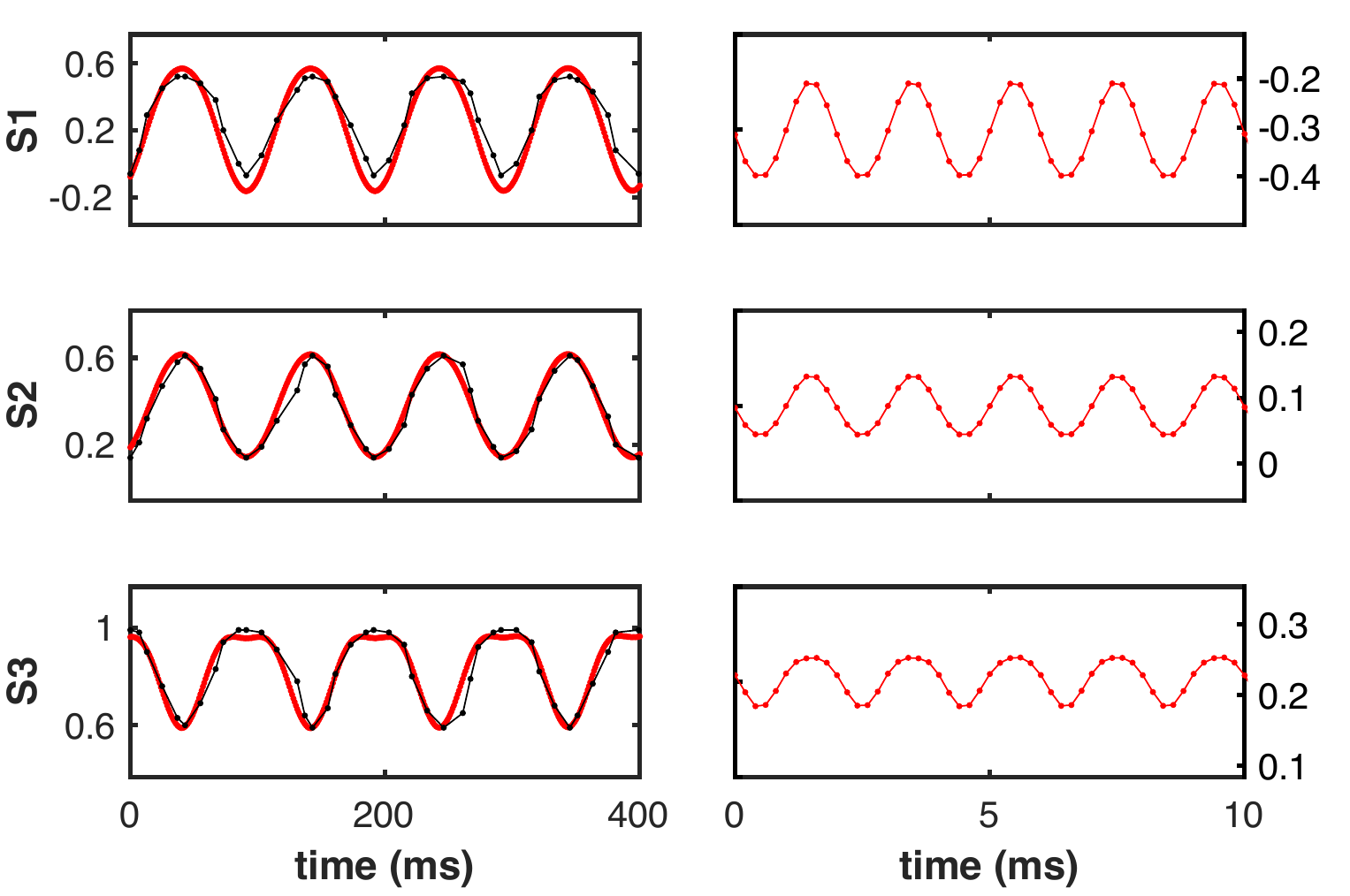}
\caption{\textbf{Speed and sampling regularity.} Right column: The Stokes parameters $S_{1}$ to $S_{3}$ as a function of time, measured by the commercial polarimeter (black) and retrieved from the speckle patterns (red), when a 10 Hz modulation is applied. The acquisition rates are respectively 150 Hz (on average) and 1000 Hz. Left column: The Stokes parameters as a function of time retrieved from the speckle patterns, when a 500 Hz modulation is applied. At this point the modulation is no longer visible on the commercial polarimeter.}
\label{fig:auto}
\end{figure}

We performed the same single-beam experiment as in section 4 but replacing the three waveplates by an electro-optic modulator (EOM, Thorlabs EO-AM-NR-C4), allowing a very rapid, electrically tunable modulation of polarisation. 
We applied a periodic modulation by applying a sinusoidal voltage to the EOM, and chose its frequency so that the commercial polarimeter made at least 10 measurements per period of modulation. Below that number of points per period, the undersampled waveform resembled random noise. As the sampling rate of the commercial polarimeter is at maximum 110 Hz, we applied a modulation of 10 Hz, leading to 11 points per period. While the polarisation state of the beam was modulated, we simultaneously recorded the measurements of the commercial polarimeter and the speckle patterns, with acquisition rates of 110 Hz and 1000 Hz respectively. We compare the measurements in figure \ref{fig:auto}. 

We further explored the speed capability by increasing the modulation frequency to 500 Hz, which is above the maximum acquisition rate of the commercial polarimeter but still clearly visible when retrieved from the speckle patterns using a camera frame rate of 5000 Hz. This was the maximum achievable frame rate in our setup before the intensity of the speckle pattern became too low. We show the corresponding measurements in figure \ref{fig:auto}. 

Although this sampling rate is about 50 times higher than that of the commercial polarimeter, it is worth pointing out that it is still much lower than what can be achieved with polarimeters based on intensity measurements after four polarising elements, sometimes called four-detector polarimeters \cite{azzam1988construction}. In that case the sampling rate is that of the photodiodes used, which can reach GHz.

\section{Summary and conclusion}
A one-to-one relation exists between the Stokes vector of a laser beam (describing its state of polarisation) and the speckle pattern it produces after diffusion. This relation takes a simple linear form, which we derived based on a linear diffusion model. We also derived an analogous relation in the case of two beams which are not coherent with each other. It involves a transmission matrix, which is unknown but can be determined through a training stage. Exploiting such a linear relation in a measurement purpose constitutes what is usually referred to as a transmission matrix method. In our case, the method essentially transfers the measurement process from the polarimeter to the camera, the knowledge being passed on during the training stage. Then the polarimeter is no longer needed and the polarisation can be retrieved from the speckle patterns alone, which unlocks several advantages. 

The main advantage we investigated is the possibility of multiplexing, i.e. measuring the polarisation state of several beams simultaneously from one single image. Another advantage we explored is the possibility of higher acquisition rate and more regular sampling than commercial polarimeters. We achieved a sampling rate of 5000 Hz, which is about 50 times higher than commercial available polarimeters. We demonstrated that the Stokes parameters of one and two beams could be retrieved with an uncertainty of 0.05, using in both cases 17 training states and 150$\times$150-pixels images, the typical resolution of a commercial polarimeter being 0.01.

Multiplexing is interesting in that it allows the embedding of information from several beams into one single image. This may find applications in optical imaging and manipulation of multiple birefringent particles. For example, in the field of levitated optomechanics, optical binding has been studied with two vaterite microparticles \cite{Arita18}. Our approach would allow the detailed analysis of the polarisation change of the light field from each particle, and hence the particle rotation and dynamics, in a facile, informative manner.

\begin{acknowledgments}
We thank Paloma Rodr\'iguez-Sevilla for technical assistance and useful discussions. We acknowledge funding from the Leverhulme Trust (RPG-2017-197) and EPSRC (EP/P030017/1).
\end{acknowledgments}

\bibliography{sample}

\end{document}